\documentclass[aps,pra,twocolumn,amsmath,amssymb,nofootinbib,showpacs,superscriptaddress,longbibliography]{revtex4-1}
\usepackage[english]{babel}
\usepackage{latexsym}
\usepackage{graphics}
\usepackage[demo]{graphicx}
\usepackage{epsfig}
\usepackage{bm}
\usepackage{amsmath}
\usepackage{amssymb}
\usepackage{amsthm}
\usepackage{dcolumn}
\usepackage{bm}
\usepackage{caption}
\usepackage{subcaption}
\usepackage{float}
\usepackage{tikz}
\usepackage{ifthen}
\usepackage{tikz-3dplot}
\usetikzlibrary{external}
\usetikzlibrary{calc}
\usepackage{pgfplots}
\pgfplotsset{compat=newest, width=8 cm}
\usetikzlibrary{arrows.meta}
\usepgfplotslibrary{fillbetween}
\usepgfplotslibrary{polar}
\usepackage{pgfplotstable}
\usepackage{hyperref}
\usepackage{comment}
\usepackage{epstopdf}
\usepackage{cancel}
\usepackage{cleveref}
\usepackage{enumerate}
\usepackage{braket}
\usepackage{lipsum}  
\usepackage[hybrid]{markdown}
\usepackage{appendix}
\usetikzlibrary{3d,angles,quotes, decorations}
\usetikzlibrary{decorations.text}
\usetikzlibrary {shapes.geometric}
\DeclareMathOperator{\tr}{Tr}

\newcommand{\bs}{\boldsymbol}

\theoremstyle{remark}

\begin{document}

\title{Finite-size effects in two-photon correlations of exciton Bose-Einstein condensates}
\author{R.D. Ivanovskikh}
\affiliation{N.L. Dukhov Research Institute of Automatics (VNIIA), Moscow 127055, Russia}
\email{roman\_ivskh@mail.ru}

\author{I.L. Kurbakov}
\affiliation{Institute for Spectroscopy RAS, Troitsk, Moscow  108840, Russia}

\author{N.A. Asriyan}
\affiliation{N.L. Dukhov Research Institute of Automatics (VNIIA), Moscow 127055, Russia}

\author{Yu. E. Lozovik}
\affiliation{Institute for Spectroscopy RAS, Troitsk, Moscow  108840, Russia}

\begin{abstract}
        Accessing two-photon statistics via Hanbury Brown and Twiss (HBT)-type measurements is essential for investigations of excitonic Bose-Einstein condensates. In this paper, we make use of quantum hydrodynamics in order to investigate the influence of finite system size on the two-photon emission intensity of a 2D condensate of excitons. We use the developed approach to calculate the two-photon decay time of exciton condensate in GaAs quantum wells and MoS$_2$ bilayers. We demonstrate that the registered signal scales on the sample size in a qualitatively different manner than the Bogoliubov theory predicts.
\end{abstract}

\maketitle
\section{Introduction}

Among the accessible experimental platforms for exploring Bose-condensation, excitons and excitonic polaritonic are, perhaps, the most attractive ones due to their strong interaction with light. This both facilitates their experimental investigations via spectroscopic measurements and implies promising applications of excitonic, polaritonic or even purely photonic condensates as coherent light sources\cite{PhysRevB.87.201301, Klaers10, maximov2023photonBosecondensatetunableterahertz}.

With optical measurements being the main experimental tool for identifying condensation and quantifying coherence of the emitted light, the key signature of condensate emergence is the multimode character of luminescence, manifested in nontrivial two-photon  correlations. As implied by the classical Hanbury Brown and Twiss (HBT) experiment \cite{BROWN1956}, photons demonstrate no bunching when being emitted from a single-mode light source (with Bose-Einstein condensate being an example) as opposed to the case of a chaotic (thermal) one.

Indeed, two-photon spectroscopy clearly distinguishes the condensed and the thermal states for both polaritonic \cite{Kasprzakg2, Tempel2012} and pure excitonic \cite{Gorbunov2009} gases, allowing experimental investigation of their phase diagrams. The increasing time resolution of two-photon spectroscopy made it possible to study dynamics of condensates. In particular, with the use of streak-cameras as photodetectors, which decreased the time-resolution up to a few ps \cite{ASSmann:10, Takemura2012}, condensation kinetics became directly observable. Such a resolution is enough to study condensate decoherence \cite{Loveg2} and even to access the relaxation dynamics of the polaritonic gas towards the BEC state \cite{Adiyatullin2015}. In addition, among the novel techniques which provide both high time resolution and high sensitivity are the ones based on frequency upconversion in a nonlinear waveguide \cite{Delteil:19}.

To describe the results of intensity correlation measurements, one has to study population build-up in a single mode, which may be qualitatively described even by the ideal Bose gas model. However, the quantitative differences due to particle interaction are significant \cite{perrin_hanbury_2012}. Moreover, especially in two dimensions, where thermal fluctuations are known to spoil long-range order \cite{Hohenberg1967, Mermin1966}, one has to carefully consider finite-size effects. In a finite two-dimensional semiconductor sample with exciton gas, one deals with a system with potentially highly depleted condensate due to both strong interactions \cite{Lozovik2007399} and diverging thermal fluctuations owing to reduced dimensionality\cite{kane_long-range_1967}. In this regime the standard Bogoliubov theory (hereafter BT) is not applicable, which motivates the current work. The aim of our study is to utilize the quantum hydrodynamic approach unified with the BT \cite{kane_long-range_1967, Voronova2018-ip, Grudinina2021} to evaluate the intensity of the two-photon radiation and to study its dependence on the system size.

The approach developed in this paper is quite general and may be applied to a variety of 2D systems, namely excitons and exciton polaritons in quantum wells as well as the ones in novel 2D materials, among which TMDCs attract much interest in the context of exciton coherence in recent years\cite{doi:10.1021/acs.nanolett.0c00633, Brunetti_2018, 10.1063/1.4900945,Tang2024}.

In the following chapters we investigate the two-photon signal emitted by a BEC of excitons in a semiconductor microcavity. We start by describing a model experimental setup to detect the two-photon signal in Section \ref{sec:exper_model}. Then follows the Section \ref{sec:micro_model} with a detailed microscopic description of the excitonic system coupled to photonic modes of the environment. In Section \ref{sec:many-body} we use techniques of many body physics in order to deduce an expression for the two-photon signal. Finally, we discuss the results with special emphasis on size-dependent features in Section \ref{sec:discussion}. We perform numerical calculations for excitons in a GaAs/AlGaAs/GaAs quantum wells and spatially indirect ones in MoS$_2$ bilayer. Moreover, we demonstrate how to apply our results to polaritonic condensates. Conclusions follow in Section \ref{sec:conclusion}.

\section{Experimental setup}\label{sec:exper_model}
For excitons in a 2D semiconductor sample of size $L\times L$, the two-photon emission intensity from unit area is as follows :
\[
P_{\rm total}=\frac{\braket{N_{1}N_{2}}}{L^2t_0},
\]
where $N_{1}$ and $N_{2}$ are the total number of photons emitted in two directions in a finite time interval $t_0$. The bracket $\braket{...}$ denotes the ensemble average:
\begin{align}
    \braket{N_1N_2}&=\sum\limits_{\bs{p}\bs{p}^{\prime}}\tr\left[\rho(t)\widehat{N}_{\bs{p}}\widehat{N}_{\bs{p}^{\prime}}\right].
\end{align}

Here the sum is over 3D momenta of emitted photons, $\rho(t)$ is the density matrix of the system under consideration (2D excitons + 3D environmental photons).

The presence of condensate manifests itself in a contribution to this sum of a sharp angularity. Namely, the one with opposite in-plane components of the emitted photons (for details, see Appendix A). Using a decomposition $\bs p =\bs q+p_z\bs e_z$ with the two terms being the in-plane and out-of-plane components of the photon momentum respectively, we may isolate this contribution ($\delta_{q, q'}$ stands for the Kronecker delta):
\begin{align}
    P_{\rm total} = \underbrace{\sum_{\bs p, \bs p^\prime}P_{\bs p, \bs p^\prime}\delta_{\bs q,-\bs q^\prime}}_{P_{\rm angled}}+\underbrace{\sum_{\bs p, \bs p^\prime}P_{\bs p, \bs p^\prime}(1-\delta_{\bs q, -\bs q^\prime})}_{P_{\rm background}}.
\end{align}

\begin{figure}[hpt]
    \includegraphics{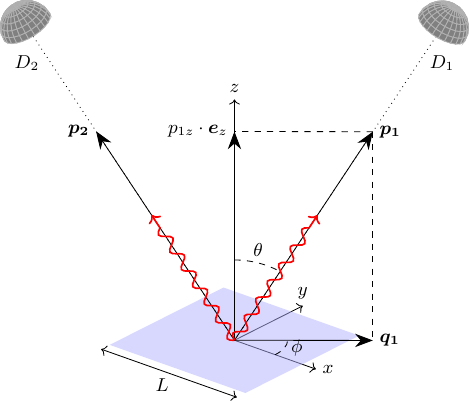}
    \caption{\raggedright A sketch of the measurement scheme for investigating the two-photon signal. Both detectors are in the same vertical plane (at angle $\phi$ in the scheme). The semiconductor sample is pictured in blue, red wavy lines denote the emitted photons.}
    \label{fig:detecting_scheme}
\end{figure}

For detection of the angled contribution, we consider an experimental setup with two photon detectors $D_1$ and $D_2$ placed above the sample as depicted on Figure \ref{fig:detecting_scheme}. The detectors are placed in the same plane, ensuring the condition  $\bs q_1=-\bs q_2$ is satisfied.

Each detector measures the number of photons emitted in a specific direction. To account for the background contribution, one may perform an additional "out-of-plane"\ measurement with a slight shift of detector positions to violate the $\bs q_1=-\bs q_2$ condition and then subtract the results of the "in-plane"\ measurement.

In order to subtract the background contribution in this manner, the detectors should be sufficiently small and far enough away to measure the far-field emission. This is essential for distinguishing a pure condensate (with macroscopic occupation of a single mode) from a quasicondensate (a bunch of low-lying states having macroscopically high occupations).

For fixed detector positions, the central quantity of our interest is the angled emission intensity from unit area in unit detector solid angle, defined as:

\begin{align}
P(\theta, \phi)\equiv\dfrac{d{P_{\rm angled}}}{d{\Omega}}
\end{align}
and the corresponding two-photon decay time, given by
\begin{align}
    \tau_2(\theta, \phi)= \frac{n}{P(\theta, \phi)},
\end{align}
where $n$ is the exciton density.

\section{The microscopic model}\label{sec:micro_model}
The two-dimensional excitonic system is described by the following Hamiltonian:
\begin{align}\label{1. H general}
\hat{H}\!=\!\sum\limits_{\bs{q}}T_{\bs q}\hat{Q}^{\dagger}_{\bs{q}}\hat{Q}_{\bs{q}}{+}\frac{1}{2L^2} \!\!\!\!\sum_{\bs{q}, \bs{q}^{\prime}, \Delta \bs{q}}\!\!\! U_0\left(\Delta \bs{q}\right) \hat{Q}_{\bs{q}}^{\dagger} \hat{Q}_{\bs{q}^{\prime}}^{\dagger} \hat{Q}_{\bs{q}^{\prime}{+}\Delta \bs{q}} \hat{Q}_{\bs{q}{-}\Delta \bs{q}}, 
\end{align}
where $\hat{Q}_{\bs{q}\lambda}$($\hat{Q}^{\dagger}_{\bs{q}\lambda}$) are the annihilation (creation) operators for excitons with momentum $\bs{q}$. The kinetic energy term is given as $T_{\bs{q}}=E_{g}+q^{2}/2m_{\text{ex}}$ with $E_{g}$ and $m_{\rm ex}$ being the semiconductor exciton bandgap and the exciton mass, respectively.

The exciton-exciton interaction potential is denoted by ${U}_0$ in (\ref{1. H general}). The findings of this study are independent of the specific form of the interaction potential, provided it ensures the stability of the uniform condensate phase. For instance, a roton instability should not be induced as it may lead to a transition to the supersolid phase \cite{PhysRevA.109.063326}. In numerical calculations for realistic expetimental setups, we will employ the dipole-dipole interaction potential for spatially separated excitons \eqref{U0(r)D2}. To address the singular behavior of the potential, we use a dressed interaction potential following \cite{Maximov2023}:
\begin{equation}\label{1. interaction}
\begin{split}
&\hat{U}=\hat{E}+\hat{U}_1, \ \ \ \ \hat{E}=\int{e}_{0}(\hat{Q}^{\dagger}(\bs{r})\hat{Q}(\bs{r}))d\textbf{r}, \\ \hat{U}_1=\dfrac{1}{2}\int&(U_{0}(\bs{r}{-}\bs{s}){-}g_{0}\delta(\bs{r}{-} \bs{s}))\hat{Q}^{\dagger}(\bs{r})\hat{Q}^{\dagger}(\bs{s})\hat{Q}(\bs{s})\hat{Q}(\bs{r})d\bs{r}d\bs{s}.
\end{split}
\end{equation}

Here, we explicitly separate the short-range contribution from the bare interaction potential $U_0(\bs r)$ by extracting a local term $g_0\delta(\bs r-\bs s)$, where $g_0={\int}U_{0}(\bs{r})d\bs{r}$. This short-range contribution is then replaced with a term that incorporates many-body effects within the local density approximation. The remaining part of the interaction is treated using the first Born approximation (see Appendix \ref{app:dressinter}). We use $e_{0}(n)$ for the free energy density component accounting for the many-body interactions \eqref{e0(n)D1}, which is extracted from the results of an {\it ab initio} numerical simulation of the 2D system of dipoles at $T=0$\cite{Lozovik2007399}. Thus, the dressed interaction is given by 
\begin{align}\label{eq:dressed_interaction}
    U(\bs r-\bs s)=U_{0}(\bs{r}-\bs{s})+\left[\frac{{\rm d}^2 e_0(n)}{{\rm d} n^2}-g_{0}\right]\delta(\bs{r}- \bs{s}).
\end{align}

To model exciton recombination, we consider a bath of 3D photons:
\begin{align}
\widehat{H}_{\text{3D}}=\sum\limits_{\bs p\lambda}\hbar\omega_{\bs p}\hat c_{\bs p\lambda}^{\dagger}\hat c_{\bs p\lambda},    
\end{align}
with $\hat{c}_{\bs p\lambda}$ and $\hat{c}
_{\bs p\lambda}^{\dagger}$ being the annihilation (creation) operators for a 3D photon with momentum $\bs p$ and polarization $\lambda$.

The exciton-photon coupling term is given as 
\begin{equation}\label{3. recombination term}
\widehat{L}_{\rm int}={\sum\limits_{\bs{p},\lambda}}\bigg(L_{\bs{p}}^{\lambda}\widehat{Q}_{\bs q}\widehat{c}_{\bs p\lambda}^{+}+L_{\bs{p}}^{\lambda*}\widehat{Q}^{\dagger}_{\bs q}\widehat{c}_{\bs p\lambda}\bigg). 
\end{equation}
Here $\bs q$ is the in-plane component of $\bs p$ in the same fashion as in Figure \ref{fig:detecting_scheme}.

The coupling constant is expressed as follows \cite{PhysRevB.53.15834}:
\begin{equation}\label{3. matrix element}
|L_{\bs{p}}^{\lambda}|^2=\dfrac{L^2}{V}\dfrac{\hbar^{2}c}{\tau_{\bs{p}}^{\lambda}\sqrt{\varepsilon}},
\end{equation}
where $V$ is the quantization volume of the photonic bath, $\varepsilon$ is the dielectric constant of the environment and $\tau_{\bs{p}}^{\lambda}$ is the radiative exciton lifetime with respect to emission of a photon into the mode $\bs p\lambda$. Considering small $\bs q$, we omit the dependence of $\tau_{\bs p}^{\lambda}$ on polarization and fix $p_z\approx E_g\sqrt{\varepsilon}/c\hbar $. Thus, the transverse momentum dependence is also further suppressed, $\tau_{\bs{p}}^{\lambda}\approx\tau_{r}$ \cite{PhysRevB.47.3832}.

\section{Two-photon emission intensity}\label{sec:many-body}
Using the microscopic Hamiltonian introduced above, one may express the desired two-photon signal in terms of the excitonic anomalous Green's function ${F}_{\bs q}(t)=-i\braket{\mathcal{T}[\widehat{Q}_{\bs{q}}(t)\widehat{Q}_{-\bs{q}}(0)]}$ (see Appendix A for details):

\begin{equation}\label{eq:general_intensity_hbt}
P(\theta, \phi)=\bigg(\dfrac{p_{0}}{2\pi\hbar\tau_{\rm r}}\bigg)^{2}\dfrac{1}{\cos\theta}\int\limits_{-\infty}^{\infty}{dt}|F_{q_0}(t)|^2.
\end{equation}

Here, we have taken into account energy conservation, which restricts photon emission to momentum $p_0=\mu\sqrt{\varepsilon}/c$ ($\mu\approx E_g$ is the condensate chemical potential). Its in-plane component magnitude is given by $q_0=p_0\sin(\theta)$. Remarkably, the anomalous Green's function is nonzero only in the presence of a Bose condensate in the system. Thus, the observation of the signal given by \eqref{eq:general_intensity_hbt} provides unambiguous evidence of the presence of excitonic BEC.

Within the framework of the standard BT, the anomalous Green's function is given by the following expression (a decay term is introduced, see Appendix \ref{damping}):
\begin{equation}\label{eq:anomalous_bogol}
\begin{split}
iF(\bs{r}, t)&=n_0-\frac1{L^2} \sum_{\bs{q} \neq 0} u_{\bs{q}} v_{\bs{q}}e^{i\bs{q}\bs{r}/\hbar} \times \\
&\times\bigg( (2n_{\bs{q}}+1)\cos\left(\dfrac{\varepsilon_{\bs{q}}t}{\hbar}\right)- i\sin\left(\dfrac{\varepsilon_{\bs{q}}|t|}{\hbar}\right)\bigg)e^{-|t|/2\tau_{\bs q}}
\end{split}
\end{equation}
and the condensate density is expressed as
\begin{align}\label{eq:hydrodynamic_conddens}
    n_0=n-\frac1{L^2}\sum_{\bs{q} \neq 0} n_{\bs{q}}-\frac1{L^2} \sum_{\bs{q} \neq 0} v_{\bs{q}}^2\left(1+2 n_{\bs{q}}\right).
\end{align}
Here the Bogoliubov transformation coefficients are introduced as
\begin{align}\label{eq:bogoluv}
    u_{\bf q}=\frac{\varepsilon_{\bf q}+T_{\bf q}}
{2\sqrt{\varepsilon_{\bf q}T_{\bf q}}},\ v_{\bf q}=\frac{\varepsilon_{\bf q}-T_{\bf q}}
{2\sqrt{\varepsilon_{\bf q}T_{\bf q}}},
\end{align}
where
\begin{align}
    \varepsilon_{\bf q}=\sqrt{T_{\bf q}\left[T_{\bf q}+2 \left(n-\frac1{L^2}\sum_{\bs{q} \neq 0} n_{\bs{q}}\right) U(\bf q)\right]}
\end{align} 
is the excitation spectrum, $n_{\bs{q}}=1/(e^{\varepsilon_{\bs{q}}/T}-1)$ --- excitation occupation number, $ U(\bf q)$ is the Fourier transform of the dressed potential \eqref{eq:dressed_interaction}.

In the expressions above one may identify the quasicondensate density
\begin{align}
    n_Q=n_0-\frac1{L^2}\sum_{\bs{q} \neq 0} n_{\bs{q}}.
\end{align}
The second term in this expression is insignificant for excitonic systems under consideration, thus we further neglect the difference between $n_{Q}$ and $n$. Note that one cannot neglect the third term in \eqref{eq:hydrodynamic_conddens} due to an infrared divergence, which we address to below.

With the help of the hydrodynamic approach along with the long-wavelength approximation, we derive a modified expression for the anomalous Green's function and the condensate fraction (see Appendix \ref{damping} for an explanation of how the decay term is introduced in the first expression):
\begin{align}\label{eq:anomalous_exponential}
&iF(\bs{r}, t)=n_{0}\exp\bigg\{-\dfrac{1}{N}\sum\limits_{\bs{q}\neq{0}}u_{\bs{q}}v_{\bs{q}}e^{i\bs{q}\bs{r}/\hbar}\\\!\!&{\times}\bigg[ (2n_{\bs{q}}{+}1)\cos\left(\dfrac{\varepsilon_{\bs{q}}t}{\hbar}\right){-} i\sin\left(\!\dfrac{\varepsilon_{\bs{q}}|t|}{\hbar}\!\right)\bigg]e^{-|t|/2\tau_{\bs{q}}}\bigg\}, \nonumber
\end{align}
\begin{align}\label{eq:conddens exponential}
\frac{n_0}n&=\exp \left(-\frac{1}{N} \sum_{\bs{q} \neq 0} v_{\bs{q}}^2\left(1+2 n_{\bs{q}}\right)\right),
\end{align}
where $N$ is the number of particles in the system. We also introduced here $\tau_{\bs{q}}$ -- the lifetime of excitation with momentum $\bs{q}$ (not to be confused with previously introduced $\tau_{\rm r}$, which is the exciton radiative lifetime). We used here the same technique as the one utilized in \cite{Voronova2018-ip} for the normal Green's function,  see also \cite{kane_long-range_1967}.

Note that the terms of both the sums in \eqref{eq:anomalous_bogol} and \eqref{eq:anomalous_exponential} are infrared divergent as $1/q^2$. However, the divergent terms do not contribute to the Fourier transform of \eqref{eq:anomalous_bogol}, which
is not the case for \eqref{eq:anomalous_exponential}. To deal with the divergence we regularize the sum utilizing a Lorentzian cutoff function as follows  (here we take into account that in BEC regime one has $\tau_c = \tau_{\bs q}\big|_{\bs q = 0}$), where $\tau_{\rm c}$ is the exciton system lifetime): 
\begin{widetext}
\begin{align}\label{eq:regularisation_with_eps}
    \ln&\left(\frac{n_0}{iF(\bs{r}, t)}\right)=\frac{1}{N}\sum_{\bs q \ne 0}\frac{mT}{q^2}\frac{1}{1+\beta^2\frac{q^2L^2}{(2\pi \hbar)^2}}e^{-|t|/2\tau_c}+\nonumber\\
    +&\int \frac{d^2\bs q}{(2\pi\hbar)^2n} \left\{u_{\bs{q}}v_{\bs{q}}e^{i\bs{q}\bs{r}/\hbar}\bigg[ (2n_{\bs{q}}+1)\cos\left(\dfrac{\varepsilon_{\bs{q}}t}{\hbar}\right)- i\sin\left(\dfrac{\varepsilon_{\bs{q}}|t|}{\hbar}\right)\bigg]e^{-|t|/2\tau_{\bs{q}}}-\frac{mT}{q^2}\frac{1}{1+\beta^2\frac{q^2L^2}{(2\pi \hbar)^2}}e^{-|t|/2\tau_c}\right\}
\end{align}
with small dimensionless $\beta$. The first term in 2D has logarithmic dependence on $\beta$:
\begin{align}
    \frac{1}N\sum_{\bs q \ne 0}\frac{mT}{q^2}\frac{1}{1{+}\beta^2q^2L^2/(2\pi \hbar)^2}=\frac{\alpha}{2\pi} \sum_{\bs n\ne 0}\frac{1}{|\bs n|^2}\frac{1}{1{+}\beta^2|\bs n|^2}=\alpha \ln\left(\frac{C_1}{\beta}\right)
\end{align}
where $\bs q=2\pi \hbar \bs n/L$ is the quantised momentum with the sum being over all 2D vectors $\bs n$ with integer coordinates. The dimensionless $\alpha$ is defined as $\alpha=T/T_0$ with $T_0=2\pi\hbar^2n/m_{\rm ex}$ being the degeneracy temperature of ideal 2D Bose gas. The coefficient $C_1$ is a shape factor of the excitonic system (pumping spot. Namely, considering a square of size $L\times L$ with periodic boundary conditions being imposed, numerical calculation leads to $C_1\approx 1.511$.

Clearly, the second term above has a logarithmic contribution due to $1/q^2$ behaviour of the integrand at small momenta, which may be cancelled by a proper choice of $\beta$ and, consequently, a cutoff momentum $q_c=\frac{2\pi \hbar}{L\beta}$. The value of $q_c$ is of the order of the smallest of momenta $\hbar/r$, $\hbar/c_st$ and $T/c_s$, after which one of the factors $e^{i\bs q \bs r/\hbar}$, $cos(\varepsilon_{\bs q}t/\hbar)$ and $n_{\bs q}$ (correspondingly) deviates from its low-momentum behaviour. For excitonic systems under consideration (namely, in GaAs quantum wells and TMDC bilayers), the smallest momentum scale is set by $\hbar/(c_s t)$ for system size $L$ in range $\approx \left[c_s\tau_{\bs q};c_s\tau_c\right]$. Thus, we denote $q_c=C_0\hbar/(c_s t)$. That is why
\begin{align}\label{eq:anomal_regularized}
        iF(\bs r&, t)=n_0\exp\Bigg({-}\alpha \ln\left(\frac{C_0 C_1  L}{2\pi c_s t}\right)e^{-|t|/2\tau_c}\nonumber\\&{-}\int \frac{d^2\bs q}{(2\pi\hbar)^2n} \left\{u_{\bs{q}}v_{\bs{q}}e^{i\bs{q}\bs{r}/\hbar}\bigg[ (2n_{\bs{q}}+1)\cos\left(\dfrac{\varepsilon_{\bs{q}}t}{\hbar}\right)- i\sin\left(\dfrac{\varepsilon_{\bs{q}}|t|}{\hbar}\right)\bigg]e^{-|t|/2\tau_{\bs{q}}}-\frac{mT}{q^2}\frac{1}{1+\frac{q^2c_s^2t^2}{C_0^2\hbar^2}}e^{-|t|/2\tau_c}\right\}\Bigg).
\end{align}
\end{widetext}
The exact value of $C_0$ is obtained by numerically integrating the second term in \eqref{eq:regularisation_with_eps} and fitting logarithmic asymptotic behavior for $\ln(q_c)\to\pm\infty$, which results in $C_0\approx 0.59\pm 0.02$ (see Appendix B).

After selecting an appropriate cutoff momentum, the second term no longer contributes logarithmically. In addition, note that $\alpha$ should be a small quantity to prevent approaching the BKT transition point, which corresponds to $\alpha_{\rm crit}\approx 0.2$ (in \cite{NelsonKost1977} $\alpha=0.25$, however, in  \cite{LOZOVIK2007487} it is demonstrated that finite-size effects on the BKT crossover as well as vortices, reduce it). Thus, one may expand up to the first order in $\alpha$ and consider the Fourier transform:
\begin{align}\label{eq:final_anomal}
    iF_{\bs q\ne \bs 0}&(t){=}{-}\frac{n_0}{n}\left(\frac{2\pi c_{s} t}{C_0C_1L}\right)^{\alpha e^{-|t|/2\tau_c}}\\
    &{\times}{u_{\bs{q}}v_{\bs{q}}\!\bigg[ (2n_{\bs{q}}+1)\cos\!\left(\!\dfrac{\varepsilon_{\bs{q}}t}{\hbar}\!\right){-} i\sin\!\left(\!\dfrac{\varepsilon_{\bs{q}}|t|}{\hbar}\!\right)\bigg]e^{-{|t|}/{2\tau_{\bs{q}}}}}.\nonumber
\end{align}

The regularizing term in the integrand of \eqref{eq:anomal_regularized} does not contribute to the $F_{\bs q}(t)$ for finite $\bs q\ne 0$ due to independence of $r$. The zeroth order term of exponent expansion is omitted for the same reason.

The Fourier transform of the BT result \eqref{eq:anomalous_bogol} is as follows:
\begin{align}
    iF_{\bs q\ne \bs 0}&(t){=}{-}{u_{\bs{q}}v_{\bs{q}}\!\bigg[ (2n_{\bs{q}}{+}1)\cos\!\left(\!\dfrac{\varepsilon_{\bs{q}}t}{\hbar}\!\right){-} i\sin\!\left(\!\dfrac{\varepsilon_{\bs{q}}|t|}{\hbar}\!\right)\bigg]e^{{-}\frac{|t|}{2\tau_{\bs{q}}}}}.
\end{align}
Compared to this expression, \eqref{eq:final_anomal} has an additional size-dependent factor, which properly describes vanishing long-range order in a uniform 2D system of infinite size. This correction is crucial for $\alpha > 0$, in agreement with \cite{Hohenberg1967, Mermin1966}, whereas for $\alpha=0$ ($T=0$), the BT works fine. In general, the lower is the condensate depletion, the smaller is the deviation of our results from the predictions of the BT.

\section{Discussion}\label{sec:discussion}
\subsection{Numerical calculations for excitons}
With the anomalous Green's function being evaluated, we have the following expression for the two-photon radiation intensity:
\begin{align}\label{eq:final_intensity_hbt}
P(\theta, \phi)&=\!\bigg(\dfrac{q_{0}}{2\pi\hbar\tau_{\rm r}}\bigg)^{2}\!\dfrac{1}{\cos\theta}\left(\dfrac{n_0}{n}\right)^2\dfrac{(n{U}(\bs q))^2}{\varepsilon_{\bs{q}}^2}\!\left(\dfrac{1}{2}+n_{\bs{q}}+n_{\bs{q}}^2\right)\nonumber\\ &\times\int\limits^{\infty}_{0}\left(\dfrac{2{\pi}c_{s}t}{C_0 C_1 L}\right)^{2\alpha\exp(- t/2\tau_c)}e^{-t/\tau_{\bs{q}}}dt.
\end{align}
This equation represents the key result of the current study. We evaluate numerically the condensate fraction $n_0/n$ and investigate the dependence of the emission intensity on the size of the system and detection angle. To further explore the implications of this result, one needs the specific form of the excitation decay time $\tau_q$ dependence on the momentum $\bs q$. We use a model expression as explained in Appendix \ref{tauq}.

Two physical realizations are considered simultaneously: excitonic gas in a GaAs/AlGaAs/GaAs quantum well and a one in a TMDC bilayer such as MoS$_{\rm 2}$/hBN/MoS$_2$. 

The GaAs electron-hole separation is considered to be $D=12$ nm wide with dielectric constant $\varepsilon = 12.5$, $m_{\rm ex}=0.22$ (in units of the free electron mass)\cite{high_spontaneous_2012}. We employ an exciton radiative decay time of $\tau_{\rm r}=20$ ns and a system decay time of $\tau_{\rm{c}}=50$ ns.

For the MoS$_{\rm 2}$/hBN/MoS$_2$ structure, we use $D=1$ nm as the distance between TMDC layers (single hBN layer is considered as a spacer), $\varepsilon = 7$, $m=0.88$ \cite{fogler_high-temperature_2014}. For this system $\tau_{\rm r}=2$ ns and $\tau_{\rm{c}}=5$ ns.

\begin{figure}[hpt]
\includegraphics{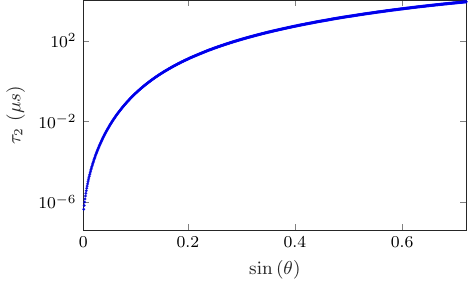}
\caption{\raggedright Angular dependence of the two-photon decay time $\tau_2(\theta, \phi)$ for exciton gas in GaAs quantum well for $n=2\times 10^{10}$ cm${}^{-2}$, $T=1$ K.}
\label{fig:angularplot}
\end{figure}

As one would expect, for increasing detection angle $\theta$ we observe a sharp decrease of emission intensity. That is to say, the two-photon decay time $\tau_2(\theta, \phi)$ increases, as depicted in Figure \ref{fig:angularplot}.
This graph does not demonstrate any qualitative deviation from the BT predictions, as well as no significant difference for TMDC excitons and quantum well excitons is revealed.

\begin{figure}[htb]
    \includegraphics[width = 0.9\linewidth]{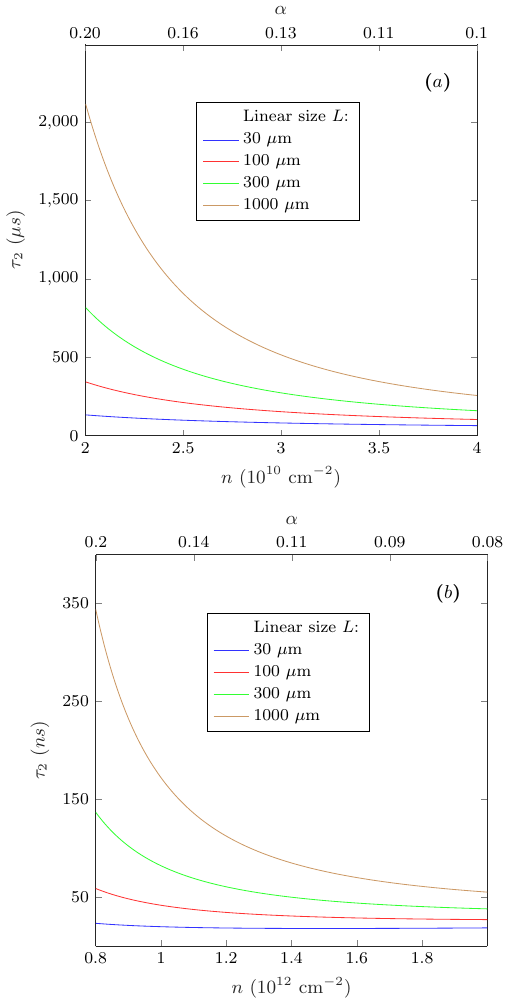}

    \caption{\raggedright The two-photon decay time $\tau_2(\theta, \phi)$ as a function of particle surface density for (a) GaAs quantum well excitons at $T=1 K$; (b) MoS$_2$ bilayer excitons at $T=10 K$; for fixed observation angle $\theta=\pi/6$. These results indicate that the increase in system size $L$ leads to a decrease in emission intensity. This is a manifestation of condensate depletion caused by thermal fluctuations.}
    \label{fig:sizedep}
\end{figure}

In contrast, we observe strong dependence on the system size for the two-photon decay time $\tau_2(\theta, \phi)$ (see Figure \ref{fig:sizedep}).

The figure highlights that the two-photon decay time is system-size dependent for fixed density. This is due to the Hohenberg-type infrared divergence (see \eqref{eq:regularisation_with_eps} and \cite{Hohenberg1967, Mermin1966}). Finite system size bounds the maximum thermal phonon wavelength. Therefore, increasing $L$ relaxes this bound, enhancing thermal fluctuations and depleting the condensate. As a result, the two-photon decay time diverges at $L\to \infty$ in agreement with the absence of Bose-Einstein condensation in a uniform infinite 2D gas.

We quantify the deviations from the BT predictions by considering the ratio:
\begin{align}\label{eq:pratio}
    \frac{P(\theta, \phi)}{P_B(\theta, \phi)}=\frac{1}{\tau_{\bs q}}\int\limits^{\infty}_{0}\left(\dfrac{2{\pi}c_{s}t}{C_0 C_1 L}\right)^{2\alpha\exp(- t/2\tau_c)}e^{-t/\tau_{\bs{q}}}dt.
\end{align}

For consistency, we use \eqref{eq:conddens exponential} even for evaluating the BT expression since expression \eqref{eq:anomalous_bogol} produces negative result for the densities of interest.

For $\tau_{\bs q}\ll\tau$, which appears to be the case for GaAs excitons observed at large enough angle $\theta$, the ratio scales as $L^{2\alpha}$, as demonstrated in Figure \ref{fig:sizedep} (a). That is clearly exlained by replacing $\exp(-t/2\tau_c)$ by unity in \eqref{eq:pratio}.

\begin{figure}[htb]
\includegraphics{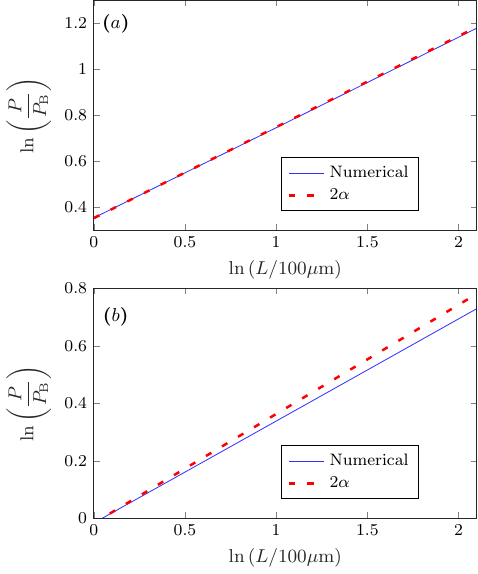}
\caption{\raggedright Scaling of the ratio $P/P_B$  with the sample linear size $L$ for (a) GaAs quantum well excitons of $n=2\times 10^{10}$ cm${}^{-2}$; (b) TMDC monolayer excitons of $n=1\times 10^{12}$ cm${}^{-2}$; for $\theta = \pi /6$. The scaling exponent is $\approx 2\alpha$ with $\alpha=0.19$ for both plots.}
\label{fig:logsizedep}
\end{figure}

For MoS${}_2$ excitons, the ratio also scales approximately with the same exponent, as shown in Figure \ref{fig:logsizedep} (b).

\subsection{Application to polaritonic condensates}
The formalism we use in this study for excitonic systems is well-applicable for exciton-polaritons also, albeit with several modifications. Namely, instead of \eqref{3. recombination term}, one should use
\begin{equation}\label{3. recombination term}
\widehat{L}={\sum\limits_{\bs{p},\lambda}}\bigg(L_{\bs{p}}^{\lambda}\widehat{c}_{\bs q\lambda}\widehat{c}_{\bs p\lambda}^{+}+L_{\bs{p}}^{*\lambda}\widehat{c}^{\dagger}_{\bs q\lambda}\widehat{c}_{\bs p\lambda}\bigg)
\end{equation}
where $\hat{P}_{\bs{q},\lambda}$ is the annihilation operator for a lower polariton, $\hat{c}_{\bs{q},\lambda}=(1-X^{2}_{\bs q , \lambda})\hat{P}_{\bs{q},\lambda}$ stands for the annihilation operator of a two-dimensional photon in the absence of upper polaritons \cite{Grudinina2021} and $X^{2}_{\bs q , \lambda}$ is the Hopfield coefficient.

In addition, one should replace the exciton lifetime $\tau_{\rm{r}}$ in the definition of $L_{\bs{p}}^{\lambda}$ by the polariton lifetime with respect to photon leakage out of the microcavity:
\begin{align}
    \frac1{\tau_{\bs q}^{\rm pol}}=\frac{1-X^2_q}{\tau_{\rm phot}}
\end{align}
with $\tau_{\rm phot}$ being the photon lifetime in the cavity. Clearly, one should also use the proper interaction potential when obtaining the condensate fraction.

However, we performed calculations for polaritons and observed only minor deviations from the BT results. This is due to small effective mass, which elevates the degeneracy temperature $T_0$, thus decreases $\alpha$, which makes the additional size scale factor negligibly different from zero. The only exception to consider is the case of extreme positive values of energy offset of the photonic spectrum with respect to excitonic one.

\section{Conclusion}\label{sec:conclusion}
In this article we present an approach to calculating the two-photon emission intensity, which properly accounts for finite-size effects. It may be used to describe the results of HBT-type measurements for condensates of excitons and exciton-polaritons in both quantum wells and novel setups with 2D materials (\textit{e.g.} TMDC layers).

By utilizing the hydrodynamic approach to the Bogoliubov theory, we considered the two-photon radiation intensity for a 2D Bose-condensate. Our results demonstrate that the standard Bogoliubov theory does not adequately predict the dependence of the radiation intensity on the size of the semiconductor sample. We claim that the modified expression we derive is the one that does. A key distinction lies in the additional scaling factor $L^{-2T/T_c}$ (where $T_c$ 
 is the degeneracy temperature for an ideal 2D gas), which effectively captures the impact of diverging thermal fluctuations that deplete the condensate at large system sizes.
 
\section{Acknowledgements}
I.L. Kurbakov and Y.E. Lozovik acknowledge the support by the Russian Science Foundation grant No. 23-42-10010, https://rscf.ru/en/project/23-42-10010/. I.L. Kurbakov acknowledges the support by the project FFUU-2024-0003 for the work described in Section IV. N.A. Asriyan acknowledges the support by the Russian Science Foundation grant No. 23-12-00115, https://rscf.ru/en/project/23-12-00115/.

\clearpage
\appendix
\section{Emission intensity}\label{sec:intensity_calc}
We start from evaluating $P_{\rm total}$, introduced in Sec. II, thus we consider the density matrix evolution:
\begin{align}
    \rho(t)=\widehat{S}(t,-\infty)\rho_{0}\widehat{S}(-\infty,t)
\end{align}
with the $S$-matrix being given by a time-ordered exponent:
\begin{align}
  \widehat{S}_{t_0}(t_2,t_1){=}\mathcal{T}\exp\bigg({-}\dfrac{i}{\hbar}\int\limits_{t_1}^{t_2}(e^{{-}\delta|t|}\widehat{U}(t){+}\theta(t)\theta(t_{0}{-}t)\widehat{L}(t))dt \bigg).
  \label{eq:time-ordered_exponent}
\end{align}

Here $\hat U$ and $\hat H_{\rm int}$ are expressed in the interaction representation with respect to the unperturbed Hamiltonian $\widehat{H}_0=\widehat{T}-\mu\widehat{N}+\widehat{H}_{3D}$ ($\hat T$ is the kinetic term). Namely,
\begin{align}
    \widehat{L}(t)={\sum\limits_{\bs{p},\lambda}}\bigg(L_{\bs{p}}^{\lambda}\widehat{Q}_{\bs q\lambda}(t)\widehat{c}_{\bs p\lambda}^{\dagger}(t)+h.c.\bigg).
\end{align}
Here $\hat c_{\bs p \lambda}^\dagger(t)=e^{i\left[\omega_{\bs p}-\mu/\hbar\right]t}\widehat{c}_{\bs p\lambda}^{\dagger}$. As in the standard diagram technique, we consider adiabatic switching of the interaction term with $t_0\delta\ll 1$, while coupling to the 3D bath is present for the time of measurement only.   

The unperturbed density matrix describes the exciton and photon subsystems separately:
\begin{align}
    \hat \rho_0\equiv\hat \rho(-\infty)=\hat \rho^{\text{exc}}{\times}\hat \rho^{\text{3D}}_0=\frac{e^{-(\widehat{T}-\mu\widehat{N})/T}}{\tr\left[e^{-(\widehat{T}-\mu\widehat{N})/T}\right]}{\times}\ket{0}\!\bra{0}_{\text{3D}},
\end{align}
with $\ket{0}_{\rm 3D}$ being the photonic vacuum.

By decomposition of the S-matrix as a product $\widehat{S}_{t_0}=\mathcal{T}\widehat{S}_{U}\widehat{S}_{Lt_{0}}$ with
\begin{equation}\label{4. W-evo}
\widehat{S}_{U}(t_2,t_1)=\mathcal{T}\exp\left(-\dfrac{i}{\hbar}\int\limits_{t_1}^{t_2}\widehat{U}(t)e^{-\delta|t|}dt \right)
\end{equation}
and
\begin{equation}\label{4. D-evo}
\widehat{S}_{Lt_{0}}=\mathcal{T}\exp\bigg(-\dfrac{i}{\hbar}\int\limits_{0}^{t_{0}}\widehat{L}(t)dt \bigg),
\end{equation} one may derive the following expression for the two-photon radiation intensity:
\begin{equation}\label{eq:2photon_signal_renormed}
P_{\rm total}=\dfrac{1}{L^2t_{0}}\sum\limits_{\bs{p}\bs{p}^{\prime}}\braket{\widehat{S}_{Lt_{0}}^{U\dagger}\widehat{c}_{\bs{p}}^{\dagger}\widehat{c}_{\bs{p}^{\prime}}^{\dagger}\widehat{c}_{\bs{p}^{\prime}}\widehat{c}_{\bs{p}}\widehat{S}^{U}_{L\tau}}.
\end{equation}

Here
\begin{equation}\label{eq:renormalized S-matrix}
\widehat{S}^{U}_{Lt_{0}}=\mathcal{T}\exp\bigg(-\dfrac{i}{\hbar}\int\limits_{0}^{t_{0}}\widehat{S}_{U}(0,t)\widehat{L}(t)\widehat{S}_{U}(t,0)dt \bigg)
\end{equation}
is the interaction-renormalized $S$-matrix, the average in (\ref{eq:2photon_signal_renormed}) is taken over the dressed density matrix 
\begin{equation}\label{eq:forward-backward_densmatrix}
\rho_H\equiv\rho(0)=\widehat{S}_{U}(0,-\infty)\rho_{0}\widehat{S}_{U}(-\infty,0).
\end{equation}

After expansion up to second order in $\widehat{L}$, we express the emission intensity
\begin{equation}\label{4. 2-phot signal expanded}
\begin{split}
&P_{\rm total}{=}\dfrac{\braket{N_{1}N_{2}}}{L^2t_{0}}{=}\dfrac{1}{L^2t_{0}}\!\sum\limits_{\bs{p}, \bs{p}^{\prime}}\!\dfrac{|L^{\lambda}_{\bs{p}}L^{\lambda}_{\bs{p}^{\prime}}|^{2}}{\hbar^{4}}\!\!\int\limits_{0}^{t_{0}}{d}t_{1}dt_{1}^{\prime}dt_{2}dt_{2}^{\prime}\\
&\times e^{- i(\hbar\omega_{\bs{p}}-\mu)(t_{1}-t_{2})/\hbar}e^{-i(\hbar\omega_{\bs{p}^{\prime}}-\mu)(t_{1}^{\prime}-t_{2}^{\prime})/\hbar}A,
\end{split}
\end{equation}
in terms of a Keldysh contour ordered product
\begin{equation}
A=\braket{\mathcal{T}_{C}[\widehat{Q}_{\bs{q}}^{U\dagger}(t_1^{\dagger})\widehat{Q}^{U\dagger}_{\bs{q}^{\prime}}(t^{\prime{+}}_{1})\widehat{Q}^{U}_{\bs{q}}(t_2^{- })\widehat{Q}^{U}_{\bs{q}^{\prime}}(t_2^{\prime{-}})]  }.
\end{equation}
The fact, that this is a time-ordered product on the Keldysh contour, is clear from the form of $\rho_H$ in \eqref{eq:forward-backward_densmatrix}. Symbols $\pm$ stand for the forward/backward branches.

Applying Wick's theorem results in four terms (subscript $_c$ stands for all connected diagrams):
\begin{align}\label{4. Wick expansion of A}
    A=&\braket{\widehat{Q}^{U\dagger}_{\bs{q}}(t_1)\widehat{Q}^{U}_{\bs{q}}(t_{2})}\braket{\widehat{Q}_{\bs{q}^{\prime}}^{U\dagger}(t_1^{\prime})\widehat{Q}^{U}_{\bs{q}^{\prime}}(t_2^{\prime})}+\nonumber\\
    &\braket{\widetilde{\mathcal{T}}[\widehat{Q}^{U\dagger}_{\bs{q}}(t_1)\widehat{Q}^{U\dagger}_{\bs{q}^{\prime}}(t_{1}^{\prime})]\mathcal{T}[\widehat{Q}^{U}_{\bs{q}}(t_2)\widehat{Q}^{U}_{\bs{q}^{\prime}}(t_2^{\prime})]}_{c}+\nonumber\\
    &\braket{\widehat{Q}^{U\dagger}_{\bs{q}}(t_1)\widehat{Q}^{U}_{\bs{q}^{\prime}}(t_{2}^{\prime})}\braket{\widehat{
Q}_{\bs{q}^{\prime}}^{U\dagger}(t_1^{\prime})\widehat{Q}^{U}_{\bs{q}}(t_2)}+\nonumber\\
&\braket{\widetilde{\mathcal{T}}[\widehat{Q}^{U\dagger}_{\bs{q}}(t_1)\widehat{Q}^{U\dagger}_{\bs{q}^{\prime}}(t_{1}^{\prime})]}\braket{\mathcal{T}[\widehat{Q}^{U}_{\bs{q}}(t_2)\widehat{Q}^{U}_{\bs{q}^{\prime}}(t_2^{\prime})]}.
\end{align}
here we used $\widetilde{\mathcal{T}}$ as a antiordering operator. Operators with superscript ${}^{U}$ are defined in the same fashion as $\widehat S^U_{L\tau}$ in \eqref{eq:renormalized S-matrix}.

Of all the terms present here, only the fourth one has the desired sharp angularity as described in Section \ref{sec:exper_model} due to a factor $\delta_{{\bs q}, -{\bs q}^{\prime}}$. Indeed, the third term is proportional to $\delta_{\bs q, \bs q'}$, thus it vanishes for spatially separated detectors as depicted in Figure \ref{fig:detecting_scheme}. The first and the second terms have smooth angular dependence (due to the assumption $\tau^{\lambda}_{\bs p}\approx\tau_{\rm r}$), thus, they contribute to background emission that is subtracted.

Substituting the fourth term from (\ref{4. Wick expansion of A}) into (\ref{4. 2-phot signal expanded}), we take the Fourier image of the anomalous Green's function $F_{\bs{q}}(t)=-i\braket{\mathcal{T}[\widehat{Q}^{U}_{\bs{q}}(t)\widehat{Q}^{U}_{-\bs{q}}(0)]}$ and consider the integrals over times $t_{1},t_{1}^{\prime},t_{2},t_{2}^{\prime}$ :
\begin{widetext}
\begin{align}
P_{\rm angled}=&\dfrac{1}{t_{0}S}\sum\limits_{\bs{p}, \bs{p}^{\prime}}\dfrac{|L^{\lambda}_{\bs{p}}L^{\lambda}_{\bs{p}^{\prime}}|^{2}}{\hbar^{4}}\bigg|\int\limits_{-\infty}^{\infty}\dfrac{d\omega}{2\pi}F_{\bs{q}}(\omega)\int\limits_{0}^{t_0}dt\int\limits_{0}^{t_0}dt^{\prime}e^{i(\hbar\omega_{\bs{p}}{-} \mu{-}\hbar\omega)t/\hbar}e^{i(\hbar\omega_{\bs{p}^{\prime}}-\mu+\hbar\omega)t^{\prime}/\hbar}\bigg|^{2}=\nonumber
\\=&\dfrac{1}{t_{0}S}\sum\limits_{\bs{p}, \bs{p}^{\prime}}\dfrac{|L^{\lambda}_{\bs{p}}L^{\lambda}_{\bs{p}^{\prime}}|^{2}}{\hbar^{4}}\bigg|\int\limits_{-\infty}^{\infty}\dfrac{d\omega}{2\pi}F_{\bs{q}}(\omega)e^{i(\hbar\omega_{\bs{p}}{-} \mu)t_{0}/2\hbar}2\pi\delta(\omega_{\bs{p}}{-} \mu/\hbar{-}\omega)e^{i(\hbar\omega_{\bs{p}^{\prime}}-\mu)t_{0}/2\hbar}t_{0}{\rm sinc}\left(\dfrac{(\hbar\omega_{\bs{p}^{\prime}}{-}\mu{+}\hbar\omega)t_{0}}{2\hbar}\right)\bigg|^{2},
\end{align}

where $\bs{p}=\{\bs{q},p_z\}$, $\bs{p}^{\prime}=\{\bs{q}^{\prime},p_z^{\prime}\}$. Hereinafter, we assume that $t_{0}$ sets the largest timescale of the problem.

Integrating over frequency $\omega$, we obtain 
\begin{equation}
P_{\rm angled}=\dfrac{1}{t_{0}S}\sum\limits_{\bs{p}, \bs{p}^{\prime}}\dfrac{|L^{\lambda}_{\bs{p}}L^{\lambda}_{\bs{p}^{\prime}}|^{2}}{\hbar^{4}}|F_{\bs q}(\omega_{\bs p}-\mu/\hbar)|^{2}t_{0}{\rm sinc}\left(\dfrac{\hbar\omega_{\bs p}+\hbar\omega_{\bs{p}^\prime}- 2\mu}{2\hbar}\right)2\pi\delta\left(\dfrac{\hbar\omega_{\bs p}+\hbar\omega_{\bs{p}^\prime}-2\mu}{\hbar}\right).
\end{equation}

\end{widetext}

Given that $F_{\bs{q}}(\omega)$ decays rapidly at $\hbar\omega>mc_{s}^{2}(\sim1\text{meV}$ in both GaAs and MoS$_2$), and $\mu\approx{E}_{g}\gg{m}c_{s}^{2}$, where $E_{g}$ is the semiconductor bandgap ($\sim{1}\text{eV}$ in both GaAs and MoS$_2$), in the sums over momenta $\bs{p}$ and $\bs{p}^{\prime}$ we can consider that $\omega_{\bs{p}}\approx\omega_{\bs{p}^\prime}\approx\mu/\hbar$. 

Thus, going from the sum over $\bs{p}$ and $\bs{p}^{\prime}$ to the integrals, given $\delta_{\bs{q},-\bs{q}^{\prime}}$, using the substitution $\omega_{1, 2}=c/(\hbar\sqrt{\varepsilon})\sqrt{q^{2}+p_{z1,2}^{2}}-\mu/\hbar$ ($p_{z1,2}>0$, the system is bounded from below by a mirror), considering $\omega_{1,2}\ll\mu$, we get
\begin{equation}
\begin{split}
P_{\rm angled}=\int\dfrac{d^2\bs{q}}{(2\pi\hbar)^{2}}\dfrac{1}{2\pi\tau_{r}^{2}}\int{\dfrac{d\omega_{1}d\omega_{2}}{1-q^{2}/p_{0}^2}}|F_{\bs{q}}(\omega_{1})|^{2}\delta(\omega_{1}{+}\omega_{2}),
\end{split}
\end{equation}
where the expression for the matrix element (\ref{3. matrix element}) was substituted and $p_0=\mu\sqrt{\varepsilon}/c$. Integrating over $\omega_{2}$ with $d\bs{q}=p_{0}^{2}\cos\theta{d}\Omega$, we obtain the expression for the two-photon signal in angular variables:
\begin{equation}
\dfrac{\Delta{P}}{\Delta\Omega}=\left(\dfrac{p_{0}}{2\pi\hbar\tau_{r}}\right)^{2}\dfrac{1}{\cos\theta}\int\limits_{-\infty}^{+\infty}\dfrac{d\omega}{2\pi}|F_{{q_0}}(\omega)|^2.
\end{equation}

\section{Anomalous Green's function calculation}\label{app_sec:anomalous}
\begin{figure}[H]
    \begin{center}
    \includegraphics{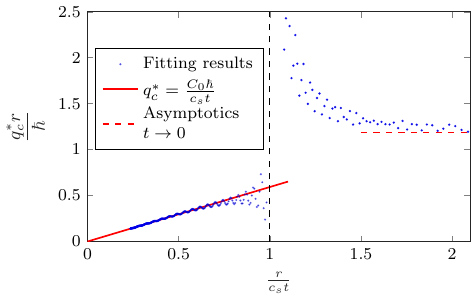}
    \end{center}
    \caption{\raggedright The dimensionless cutoff momentum $rq^*_c/\hbar$ as a function of the ratio $r/c_s t$. The linear part of the graph is utilized to derive the slope $C_0=0.59\pm 0.02$.}
    \label{fig:c0estim}
\end{figure}

As described in the main text, the expression for the anomalous Green's function \eqref{eq:anomal_regularized} has a contribution, which is logarithmically dependent on the system size. To properly extract it, we evaluate numerically the second term in the exponent, which is as follows:
\begin{align}\label{eq:fit_func}
    &I(q_c) = \!\int \!\! \frac{d^2\bs q}{(2\pi\hbar)^2n} \left\{{-}\frac{mT}{q^2}\frac{1}{1{+}\frac{q^2}{q_c^2}}e^{{-}|t|/2\tau_c}\right.\nonumber\\    &\left.+u_{\bs{q}}v_{\bs{q}}e^{i\bs{q}\bs{r}/\hbar}\bigg[ (2n_{\bs{q}}{+}1)\!\cos\!\left(\!\dfrac{\varepsilon_{\bs{q}}t}{\hbar}\!\right){-}i\!\sin\!\left(\!\dfrac{\varepsilon_{\bs{q}}|t|}{\hbar}\!\right)\bigg]e^{{-}|t|/2\tau_{\bs{q}}}\right\}=\nonumber\\
    &=\alpha \ln\left(\frac{q^*_c}{q_c}\right)e^{-|t|/2\tau_c}+\Delta I(q_c),
\end{align}
for different values of $q_c$. Then, we fit asymptotic behavior of $I(q_c)$ at $q_c\to \pm \infty$, where logarithmic contribution is dominant with a function $f(q_c)=\alpha \ln\left(q^*_c/q_c\right)\exp(-|t|/2\tau_q)$. From the fitting, we derive the value of cutoff momentum $q_c^*$, which cancels the logarithmic contribution. This procedure is implemented for various values of $t$ for fixed $r$ and below we present the results in Figure \ref{fig:c0estim}, namely the fitted value $rq^*_c/\hbar$ as a function of $r/c_st$.
From the linear region of the graph, we derive the desired value of the slope $C_0=0.59\pm 0.02$.

\section{Introducing damping into the anomalous Green's function}\label{damping}
Considering the number of excitons $N$ to be finite and their decay to be weak ($\tau_c\gg L/c_s$,
\cite{prl099140402}), one may neglect the difference between the superfluid $n_s$, the locally
superfluid $n_l$, the quasi-condensate $n_Q$ (see Sec. \ref{sec:many-body}) density and the total density $n$ in a sufficiently pure spatially homogeneous strongly correlated \cite{Lozovik2007399} exciton system at $\alpha\lesssim0.2$. Thus, we apply the formalism developed in \cite{Voronova2018-ip}
for normal correlators to the calculation of the anomalous Green's function:
\begin{align}\label{FrtC1}
iF({\bf r},t)&=n\exp(-\langle\mathcal{T}[
(\hat\varphi({\bf r},t)+\hat\varphi(0,0))^2/2]\rangle)=\nonumber\\
&=n_0\exp(-\langle\mathcal{T}[\hat\varphi({\bf r},t)\hat\varphi(0,0)]\rangle).
\end{align}

Here $n_0=n\exp\left(-\braket{\hat\varphi^2(0,0)}\right)$ is the density of the condensate (\ref{eq:conddens exponential}) and
\begin{equation}\label{phirtC2}
\hat\varphi({\bf r},t)=\frac i{\sqrt N}\sum_{{\bf q}\ne0}
\sqrt{\frac{\varepsilon_{\bf q}}{4T_{\bf q}}}(\hat\alpha_{\bf q}(t)-\hat\alpha^+_{\bf -q}(t))e^{i{\bf qr}/\hbar}
\end{equation}
is the Heisenberg phase operator with $\hat\alpha_{\bf q}(t)$ being the bosonic excitation annihilation operator with momentum ${\bf q}$.

Substituting (\ref{phirtC2}) into (\ref{FrtC1}) and taking into account the absence of anomalous
averages for the excitations, \textit{i.e.}
$\langle\hat\alpha_{\bf q}(t)\hat\alpha_{\bf q'}(t')\rangle=0$, we deduce
\begin{align}\label{FrtC3}
iF({\bf r},t)&=n_0\exp\!\Big[-\!\sum_{{\bf q}\ne0}\!
\frac{e^{i{\bf qr}/\hbar}}Nu_{\bf q}v_{\bf q}\nonumber\\
&\times\langle\mathcal{T}[\hat\alpha_{\bf q}(t)\hat\alpha^+_{\bf q}(0){+}
\hat\alpha^+_{\bf q}(t)\hat\alpha_{\bf q}(0)]\rangle\!\Big]\!,
\end{align}
where we used $u_{\bf q}v_{\bf q}=n{U}(\bs q)/2\varepsilon_{\bs q}$ (see \eqref{eq:bogoluv}). We further substitute an ultraviolet cutoff factor $\varkappa_{\bf q}=1-T_{\bf q}^2/\varepsilon_{\bf q}^2$  in \eqref{FrtC3}, following \cite{Voronova2018-ip}. This specific form ensures a proper unification of the quantum-field hydrodynamics with Bogoliubov's theory for anomalous correlators.

The sum of time correlators in \eqref{FrtC3} in the inner brackets
of the excitation operators can be expressed through their retarded $\mathcal G^R_{\bf q}(\omega)$ and the advanced $\mathcal G^A_{\bf q}(\omega)$
Green's functions as
\begin{align}\label{C4}
\langle{\mathcal T}[&\hat\alpha_{\bf q}(t)\hat\alpha^+_{\bf q}(0){+}
\hat\alpha^+_{\bf q}(t)\hat\alpha_{\bf q}(0)]\rangle=\\=&\int_{-\infty}^{\infty}
\frac{d\omega}{2\pi}[i\mathcal G_{\bf q}(\omega){+}
i\mathcal G_{\bf q}(-\omega)]e^{-i\omega t},
\vspace{-0mm}
\end{align}
where the causal excitation Green's function is \cite{AGD}
\vspace{-2mm}
\begin{equation}\label{GqoC5}
\mathcal G_{\bf q}(\omega)=(1+N(\omega))\mathcal G_{\bf q}^R(\omega)-
N(\omega)\mathcal G_{\bf q}^A(\omega),
\vspace{-2mm}
\end{equation}
with $N(\omega)=1/(e^{\hbar\omega/T}-1)$.

In turn, the retarded (advanced) Green's function is given by the analytic continuation of the Matsubara Green's function $\mathcal G_{\bf q}^M(\omega_s)$ to the upper (lower) half-plane \cite{AGD}
\begin{equation}\label{GRAqoC6}
\mathcal G_{\bf q}^{R(A)}(i\omega_s)=\mathcal G_{\bf q}^M(\omega_s),\;\;\;\;\;
\omega_s>0\;(<0).
\vspace{-2mm}
\end{equation}
Here $\omega_s=(2\pi T/\hbar)s$ is the Matsubara frequency and $s$ is an integer.

In a system with damping, the Matsubara Green's function has the standard form \cite{Mahan}
\vspace{-2mm}
\begin{equation}\label{GMqoC7}
\mathcal G_{\bf q}^M(\omega_s)=\frac1{i\omega_s-\varepsilon_{\bf q}/\hbar+
i\rm{ sign}(\omega_s)/(2\tau_{\bf q})},\;\;\;\;\omega_s\ne0.
\vspace{-2mm}
\end{equation}

Sequentially substituting \eqref{GMqoC7} into \eqref{GRAqoC6}, \eqref{GRAqoC6} into \eqref{GqoC5}, \eqref{GqoC5} into \eqref{C4}, and \eqref{C4} into (\ref{FrtC3}), we obtain the final
expression (\ref{eq:anomalous_exponential}) for the anomalous Green's function.

A similar calculation in the framework of the standard Bogoliubov theory is based on the
expressions (\ref{GqoC5})-(\ref{GMqoC7}) and leads to (\ref{eq:anomalous_bogol}).

\section{Dressed interaction}\label{app:dressinter}
Here, we specify the form of ${U}({\bf q})$ (\ref{eq:dressed_interaction}) that is used in our calculations. We proceed in the same fashion as in our previous study \cite{PhysRevB.95.245430}. Namely, as explained in the main text, the dressed coupling constant $g$ is extracted from the results of an \textit{ab initio} simulation \cite{Lozovik2007399}:
\begin{align}
    g\equiv U(0)=\frac{{\rm d}^2e_0(n)}{{\rm d} n^2}
\end{align}
with
\begin{equation}\label{e0(n)D1}
e_0(n)\!=\!\frac{d^2}{\varepsilon r_D^5}a_1
\exp[(1+a_2)\ln u+a_3\ln^2u+a_4\ln^3u+a_5\ln^4u].
\end{equation}

Here $r_D=m_{\rm ex}d^2/\hbar^2\varepsilon$ and $u=nr_D^2$ is the dimensionless density,
$d$ is the dipole moment of the exciton, $\varepsilon$ stands for the dielectric constant of the surrounding medium, and the analytical fit to the numerical
simulation results is for $0.004\le u\le8$ with coefficients being
$a_1=9.218$,
$a_2=1.35999$, $a_3=0.011225$, $a_4=-0.00036$, and $a_5=-0.0000281$.

We compose the dressed interaction of the dressed coupling constant $g$ and the bare remnant:
\begin{align}
     U(\bs r) = g\delta(\bs r) + \left[U_0(\bs r)-\delta(\bs r)\int {\rm d}^2\bs r' U_0(\bs r')\right].
\end{align}
For the latter we utilize the interaction potential for "separated dipoles"\ , which is as follows:
\begin{equation}\label{U0(r)D2}
U_0(r)=\frac{2e^2}{\varepsilon}\left(\frac1r-\frac{1}{\sqrt{r^2+D^2}}\right),
\end{equation}
where $e$ is the elementary charge and $D=d/e$ is the electron-hole separation.

\section{Introducing the excitation lifetime}\label{tauq}
Calculating the excitation decay times for excitons is a complex problem due to multiple contributing decay channels: scattering on lattice phonons, free carriers, disorder potential, \textit{etc.}

Not to deal with all these details, which do not affect the qualitative predictions of our study, we use the following model expression for the excitation decay rate:
\begin{equation}\label{tauqD3}
\frac{\hbar}{\tau_{\bf q}}=\frac{\hbar}{\tau_c}+\Gamma_{\bf q},\;\;\;\;
\Gamma_{\bf q}=\gamma_0T_{\bf q}\exp(-T_{\bf q}/gn).
\end{equation}
Such a choice is motivated by the following simplistic considerations:
\begin{enumerate}
    \item The damping $\hbar/\tau_c$ associated with the decay of the system is added to the $\Gamma_{\bf q}$ additively.
    \item The contribution of $\Gamma_{\bf q}$ at small momenta is quadratic. Omission of the linear term is motivated by the fact that the thermal channel \cite{popov_functional_1983} contribution to the damping $\Gamma_{\bf q}$ is linearly dependent on $q$ and scales as a high power of $T$. The latter is low ($T\ll gn,T_0$), so this channel is weak.

    Other damping channels contribute with terms scaling as a power of momentum, which is higher than unity, \textit{e.g.} the white-noise disorder contribution, as demonstrated in \cite{giorgini_effects_1994}. 
  
    \item For high momenta, $\Gamma_{\bf q}$ is a decreasing function.
\end{enumerate}

In numerical calculations we use $\gamma_0=0.1$, which is to set the contribution of $\Gamma_{\bf q}$ to the order of a tenth of the value of the excitation spectrum itself for moderate momenta (one could choose any small value for which condensate still exists, this choice does not affect the results qualitatively).
\clearpage
\bibliography{_main}
\end{document}